\documentclass[aps,prb,superscriptaddress,showpacs,twocolumn]{revtex4}
\usepackage{graphicx}
\usepackage{amssymb}
\usepackage{amsmath}
\usepackage{color}
\usepackage{bm}

\newcommand{\sip}{\sigma_{\rm p}}
\newcommand{\etap}{\eta_{\rm p}^{\rm r}}
\newcommand{\zp}{z_{\rm p}}
\newcommand{\pco}{p_{\rm co}}
\newcommand{\un}{\frac{k_{\rm B}T}{\sigma^3}}
\newcommand{\una}{\frac{k_{\rm B}T}{\sigma^2}}
\newcommand{\tg}{\tilde{\gamma}}

\begin{document}

\title{Monte Carlo simulations of the solid-liquid transition in hard 
spheres and colloid-polymer mixtures}
\author{T. Zykova-Timan$^{1}$, J. Horbach$^{2}$, and K. Binder}
\affiliation{Institut f\"ur Physik, Johannes 
Gutenberg-Universit\"at Mainz, Staudinger Weg 7, 55099 Mainz, Germany\\
$^{\it 2}$Institut f\"ur Materialphysik im Weltraum, Deutsches Zentrum 
f\"ur Luft- und Raumfahrt (DLR), 51170 K\"oln, Germany}

\date{\today}

\begin{abstract}  
Monte Carlo simulations at constant pressure are performed to
study coexistence and interfacial properties of the liquid-solid
transition in hard spheres and in colloid-polymer mixtures. The
latter system is described as a one-component Asakura-Oosawa (AO)
model where the polymer's degrees of freedom are incorporated via an
attractive part in the effective potential for the colloid-colloid
interactions. For the considered AO model, the polymer reservoir
packing fraction is $\etap=0.1$ and the colloid-polymer size ratio
is $q\equiv\sip/\sigma=0.15$ (with $\sip$ and $\sigma$ the diameter
of polymers and colloids, respectively). Inhomogeneous solid-liquid
systems are prepared by placing the solid fcc phase in the middle of a
rectangular simulation box creating two interfaces with the adjoined
bulk liquid. By analyzing the growth of the crystalline region at
various pressures and for different system sizes, the coexistence
pressure $\pco$ is obtained, yielding $\pco=11.576\;\un$ for the hard
sphere system and $\pco=8.0\;\un$ for the AO model (with $k_{\rm B}$ the
Boltzmann constant and $T$ the temperature). Several order parameters are
introduced to distinguish between solid and liquid phases and to describe
the interfacial properties. From the capillary-wave broadening of the
solid-liquid interface, the interfacial stiffness is obtained for the
$(100)$ crystalline plane, giving the values $\tg \approx 0.49\;\una$
for the hard-sphere system and $\tg\approx 0.95\;\una$ for the AO model.
\end{abstract}

\maketitle

\section{Introduction}
\label{sec:intro} 
Various colloidal systems are ideal models for the investigation
of crystal nucleation and crystal growth processes. Whereas in
atomistic systems, nucleation rates or interfacial free energies
are hardly experimentally accessible from direct measurements,
in colloidal systems the much larger length and time scales
allow to determine these properties, at least in principle. For
instance, \emph{in situ} measurements of static structure factors
in hard sphere-like colloidal systems using light scattering
techniques resulted in estimates of nucleation rates and have
given insight into the applicability of classical nucleation theory
\cite{schaetzel92,harland97,schoepe06}. Moreover, confocal microscopy
gives a direct access to particle trajectories and thus, similar as in
a computer simulation, any quantity of interest can be computed from
the positions of the particles. Recently, several experimental studies
using confocal microscopy \cite{gasser01,dullens06,prasad07,guzman09}
have revealed various properties of solid-liquid interfaces.  However,
a direct estimate of anisotropic interfacial free energies in colloidal
systems has not been possible so far.

Two paradigms of colloidal model systems that can be realized
experimentally are hard spheres and hard spheres with a short-ranged
attractive interaction (colloid-polymer mixtures). Due to the short-range
of the interactions, these model systems are also well-suited for
theoretical studies (e.g.~in the framework of density functional
theory \cite{curtin88,kahl09}) and for computer simulations. As far
as the solid-fluid transition in hard spheres is concerned, various
Molecular Dynamics (MD) and Monte Carlo (MC) techniques have been used
to estimate thermodynamic properties such as the coexistence pressure
\cite{frenkel88,polson98,wilding00,errington04,vega07}, kinetic
growth coefficients \cite{amini06}, and interfacial free energies
\cite{laird00,laird06,tanya09}. For the Asakura-Oosawa (AO) model of
colloid-polymer mixtures (see below), MC studies \cite{dij99,dij06}
have provided estimates for the solid-liquid phase boundaries in a wide
range of model parameters. However, to our knowledge, interfacial free
energies for solid-fluid interfaces have not been determined so far for
the AO model.

Despite the efforts that have been recently undertaken, the examination
of solid-liquid interfaces even for hard-sphere and hard-sphere-like
systems is still subject to various problems. In particular, the role of
finite-size effects has not been investigated in a systematic manner.
Recently \cite{tanya09}, we have made a first preliminary step to
fill this gap considering the solid-liquid interfaces of hard spheres
and of the metallic system Ni. In the latter work, we have estimated
the coexistence pressure and the interfacial stiffness $\tilde{\gamma}$
(see below) using constant-pressure MC simulations of solid-liquid
inhomogeneous systems.  To estimate $\tilde{\gamma}$, a result of
capillary wave theory (CWT) was employed, according to which for a rough
interface the mean-squared width of the interface grows logarithmically
with the lateral size of the system. Thus, we have made use of finite-size
effects to compute the interfacial stiffness $\tilde{\gamma}$.  

In the present work, much more extensive calculations are performed to
determine the coexistence pressure and various interfacial properties. In
addition to the hard-sphere system, the AO model for colloid-polymer
mixtures is considered. As a matter of fact, it is much more difficult
to compute interfacial properties for the latter colloid-polymer model
than for the hard sphere system due to a slower growth kinetics as well
as smaller-amplitude capillary fluctuations along the interface. However,
we show that both for the hard sphere system and the AO model accurate
values for the coexistence pressure are obtained without the requirement
of extrapolation from relatively small system sizes to the thermodynamic
limit.

The rest of the paper is organized as follows. In the next section
the interaction models are introduced and the main details of the
simulation are given. The results on the solid-fluid coexistence pressure
are worked out in Sec.~\ref{sec:coex}, followed by the introduction of
local order parameters and the discussion of the interfacial structure
in Sec.~\ref{sec:order}. Then, Sec.~\ref{sec:cwm} is devoted to
the determination of the interfacial stiffness from the finite-size
broadening of the interface.  Finally, a summary and discussion of the
results is provided in Sec.~\ref{sec:conclusions}.

\section{Model systems and simulation techniques}
\label{sec:model}
The one-component hard sphere model can be considered as the simplest
model with a solid-liquid transition. For a system of hard spheres of
diameter $\sigma$, the interaction potential is defined by
\begin{equation}
\label{eq:hsmodel}
V_{\rm HS}(r) = \left\{
\begin{array}{ll}
\infty & r<\sigma \\
0      & r\geq\sigma \, ,
\end{array}
\right.
\end{equation}
with $r$ the distance between two particles. The freezing of hard
spheres has been first observed in early molecular dynamics simulations
\cite{alder57,wood57,hoover68}. In these simulations, systems of $N=500$
particles were considered. In this work, systems of up to $10^5$ particles
are investigated; in particular to obtain a reliable estimate of the
interfacial stiffness $\tg$.

To a very good approximation, the equation of state for the fluid phase
of the hard sphere model is given by the analytical Carnahan-Starling
equation \cite{cs69,hansen} and to a lesser degree for the solid by
the Hall equation \cite{hall72}. Accurate values for the thermodynamic
properties of the bulk solid are provided by empirical fits to computer
simulations \cite{speedy98}. Since any allowed hard-sphere configuration
has zero potential energy, the solid-fluid transition in the hard-sphere
system is completely driven by entropy and temperature $T$ plays the role
of a scaling factor. As a result, the thermodynamic properties are fully
controlled by the packing density $\eta=\frac{\pi\sigma^3}{6}\frac{N}{V}$
(or, by the pressure $p$ in case of fluctuating total volume $V$). At
first glance, it seems to be surprising that hard spheres solidify since
one may expect the entropy of an ordered solid phase to be always lower
than that of the disordered fluid phase. However, at sufficiently high
packing fractions, the spheres in a solid fcc configuration have locally
more freedom to move than in a fluid at the same packing fraction, and
the resulting higher number of possible microstates for the solid phase
corresponds to a higher entropy.

Experimentally, the repulsive interactions between hard sphere colloids
can be modified by the addition of non-adsorbing polymers. Although the
pairwise colloid-polymer as well as the polymer-polymer interactions
are repulsive, an effective attraction between the colloids in
colloid-polymer mixtures is induced entropically by a depletion effect
\cite{gast83,fvt92}.

\begin{figure}
\begin{center}
\includegraphics[width=0.4\textwidth,clip]{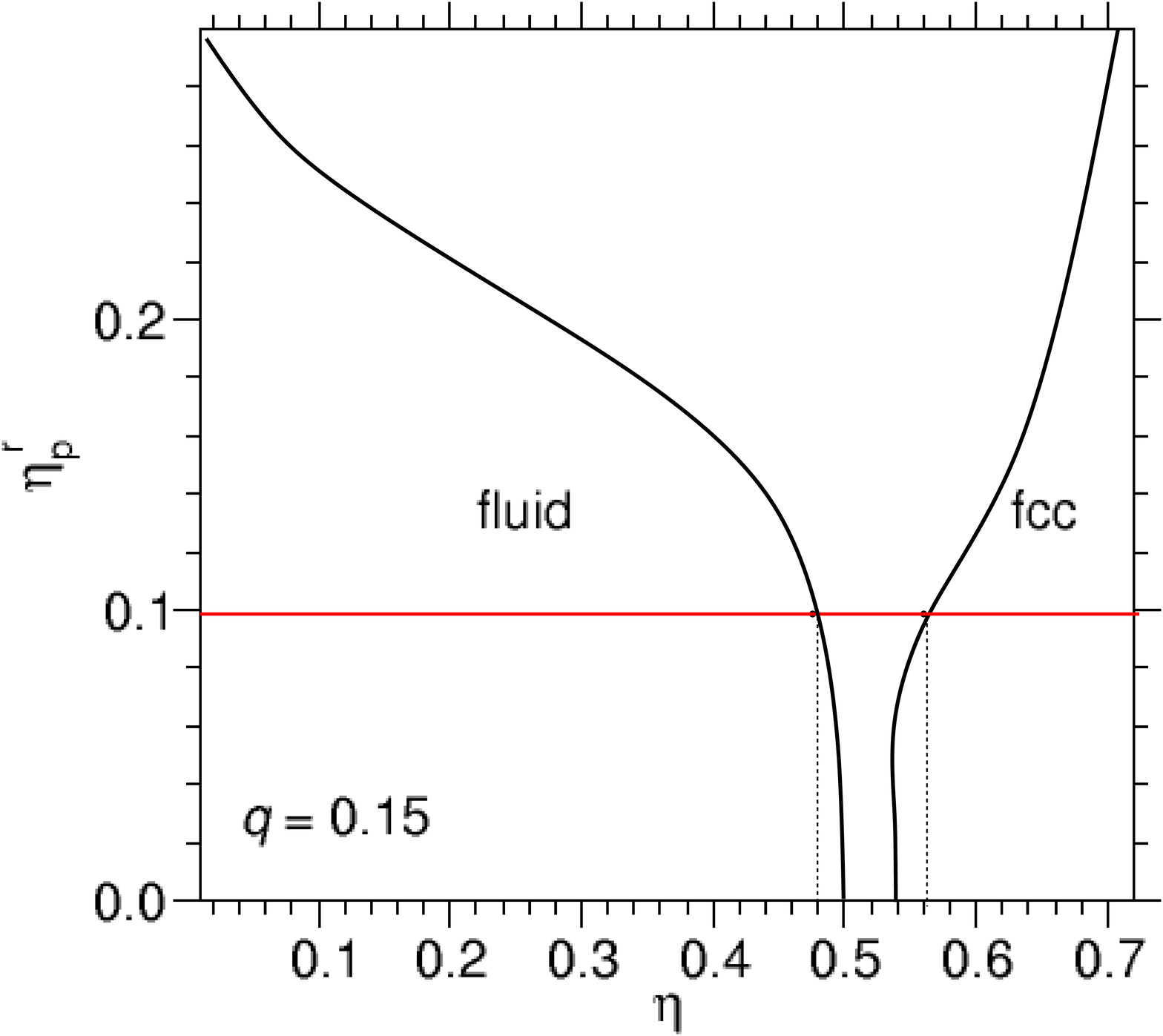}
\caption{A sketch of the phase diagram for the colloid-polymer mixture in 
the $\etap-\eta$ plane, assuming a polymer-colloid size ratio $q=0.15$.
Throughout this work, we consider either $\etap=0.0$ (hard sphere model) 
or $\etap=0.1$ (marked by the horizontal line).}
\label{fig1}
\end{center}
\end{figure}
A simple model for colloid-polymer mixtures is the
Asakura-Oosawa (AO) model \cite{ao54,ao58,vrij76} where a hard
sphere interaction is assumed between colloids as well as between
colloids and polymers while polymer particles do not interact
with each other.  The limits of this approximation are discussed
elsewhere~\cite{meijer94,flory}.  In this two-component model,
the strength of the attractive interactions between the colloids is
controlled by the density of polymer particles and the range of attraction
by the size ratio between polymers and colloids, $q\equiv \sip/\sigma$
(with $\sip$ and $\sigma$ corresponding to the diameter of polymers and
colloids, respectively).  If one considers the system to be coupled to
a polymer reservoir, the fugacity of polymers $\zp$ (or the polymer
reservoir packing fraction $\etap\equiv \frac{\pi \sip^3 \zp}{6}$)
can be regarded as the analog of inverse temperature in a molecular
system and the phase diagram can be displayed in the $\etap-\eta$
plane (corresponding to the temperature-density plane in a molecular
system). At $\etap=0$ the coexistence region reduces to the case of
pure hard spheres, whereas the increase of the polymer fugacity $\zp$
broadens the coexistence region, corresponding to a higher coexistence
packing fraction of the solid phase and a lower one of the fluid phase
(Fig.~\ref{fig1}). Note that throughout this work the polymer reservoir
packing fraction is fixed at $\etap=0.1$, as indicated by the horizontal
line in Fig.~\ref{fig1}.

On a qualitative level the AO model provides a good description
of a colloid-polymer mixture. Modifications of this model that
lead to a more reliable description have been proposed elsewhere
\cite{meijer94,vink05_2,zausch09,zausch10}. On the other hand, the AO
model can be simplified in terms of the computational load by integrating
out the polymers degrees of freedom and represent the colloid-polymer
mixture as a one-component system of colloidal particles that interact
with each other via an effective interaction potential. While for
$q\ge 0.154$, this potential is a sum of two-body, three-body and
higher-body terms, for $q<0.154$ it reduces to a pair potential given
by \cite{ao54,ao58,dij02}
\begin{equation}
\beta V_{\rm AO}(r)= 
\begin{cases}
-\eta_{\rm p}^{\rm r}\frac{(1+q)^3}{q^3} \times & \\
\left( 
1 -\frac{3r/\sigma}{2(1+q)}+
\frac{(r/\sigma)^3}{2(1+q)^3}\right) 
& \sigma<r<\sigma+\sigma_{\rm p} \\ 
0 & r>\sigma+\sigma_{\rm p}       
\end{cases}
\label{eq_vao}
\end{equation}
with $r$ the distance between two colloids and $\beta\equiv (k_{\rm
B}T)^{-1}$ (with $k_{\rm B}$ the Boltzmann constant).

In this work, the fluid-to-solid transition of the hard sphere system and
the AO model, as described by the potential (\ref{eq_vao}), is studied
using Monte Carlo (MC) simulations in the isothermal-isobaric $NpT$
and $Np_zT$ ensembles, i.e.~at constant pressure $p$, temperature $T$
and particle number $N$ (in the $Np_zT$ ensemble, $p_z$ is the diagonal
component of the pressure tensor perpendicular to the $xy$ plane, 
i.e.~perpendicular to the interface). MC
simulations were carried out using a standard Metropolis algorithm.
The trial moves were particle displacements and a rescaling of the volume
for one MC cycle \cite{binder-book,frenkel-book,krauth-book}. The maximum
particle displacement was chosen such that the acceptance rate maintained
constant at 30\% for particle displacements and at 10\% for volume's
rescaling. To optimize the speed of the simulation for large system sizes,
a cell-linked neighbor list was used \cite{frenkel-book,krauth-book}.

\section{Results}
\begin{figure}[h]
\begin{center}
\includegraphics[width=0.4\textwidth,clip]{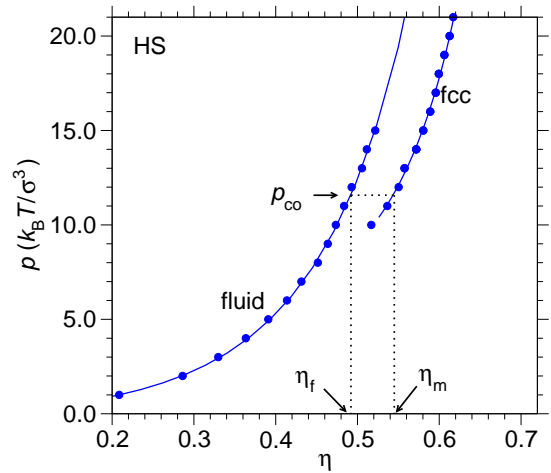}
\caption{Phase diagram of the hard sphere system in the $(p,\eta)$
plane. The filled circles are simulation results, whereas the solid
lines correspond to analytical expressions estimates of the equation of
state, as proposed by Carnahan and Starling (fluid branch) and by Hall
(solid branch).  $p_{\rm co}\approx11.576$ is the estimated coexistence
pressure at which the transition from a fluid to a crystalline fcc phase
occurs. The coexistence region is between the freezing point at $\eta_{\rm
f}\approx 0.492$ and the melting point at $\eta_{\rm m}\approx 0.545$.}
\label{fig2}
\end{center}
\end{figure}
\begin{figure}[h]
\begin{center}
\includegraphics[width=0.4\textwidth,clip]{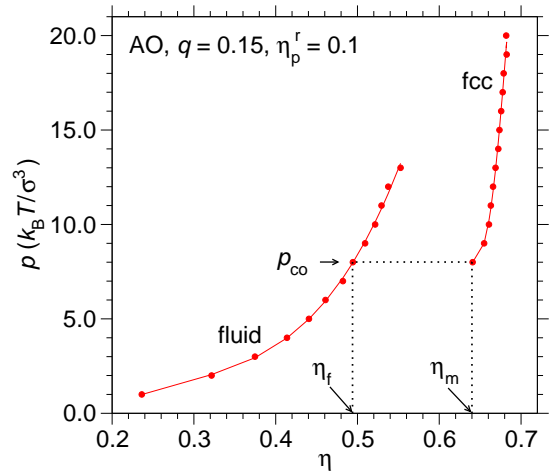}
\caption{Phase diagram of the AO model in the $(p,\eta)$ plane for
$q=0.15$ and $\eta_{\rm p}^{\rm r}=0.1$. The filled circles are the simulation
results. Here, the solid lines are spline interpolations between
simulation points. The estimated coexistence pressure is $p_{\rm
co}\approx8.0$. Freezing and melting points are at $\eta_{\rm f}\approx
0.494$ and $\eta_{\rm m}\approx 0.64$, respectively.}
\label{fig3}
\end{center}
\end{figure}
\begin{figure}[h]
\begin{center}
\includegraphics[width=0.4\textwidth,clip]{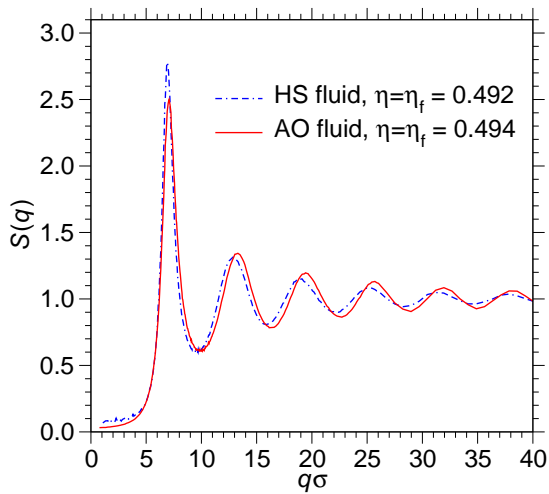}
\caption{Static structure factors $S(q)$ of the hard sphere and the
AO fluid at the freezing point.}
\label{fig4}
\end{center}
\end{figure}
\subsection{Coexistence pressure}
\label{sec:coex}
As a prerequisite for the investigation of interfacial properties,
an accurate determination of the coexistence pressure $p_{\rm co}$
is required. To this end, inhomogeneous systems are prepared where the
crystal phase in the middle of an elongated simulation box is surrounded
by the fluid phase, separated by two planar interfaces (note that two
interfaces appear due to the use of periodic boundary conditions).
When solid and liquid are in equilibrium at the pressure $p_{\rm co}$,
the thermodynamic driving force is zero and the average total volume
$\langle V(t) \rangle$ of the system does not change as a function of
time, i.e.~the crystal neither grows nor melts. In this manner, $p_{\rm
co}$ can be identified as the pressure where the time derivative $\langle
dV(t)/dt \rangle$ vanishes.

To prepare an inhomogeneous system at a given pressure, one has to first
compute separately the equation of state, $p(\eta)$, for the pure fluid
and the pure crystal.  From the fluid and the solid branch of $p(\eta)$
one can then read off the packing fraction $\eta$ (or the volume $V$)
of the two phases at a given pressure. For the calculation of $p(\eta)$,
we used MC simulations in the $NpT$ ensemble. These runs were done
for systems of $N=500$ and, in some cases, for $N=1728$ particles.
The simulations with the larger system size indicated that finite-size
effects are negligible for the calculation of the equation of state, at least
for systems with $N\ge 500$ particles. The simulation time  
was chosen depending on the convergence of the results, ranging
from 500000 to several million MC cycles.

Figures~\ref{fig2} and \ref{fig3} display respectively the equation
of state of the hard sphere and the AO systems (with $q=0.15$ and
$\etap=0.1$). For the hard sphere system, the simulation results are
well-described by analytical expressions (solid lines), as proposed by
Carnahan and Starling \cite{cs69} for the liquid branch and by Hall
\cite{hall72} for the fcc phase.  In the case of the AO system, no
accurate analytical predictions are available and so the solid lines
in Fig.~\ref{fig3} are just spline interpolations that connect the
data points. Also indicated in Figs.~\ref{fig2} and \ref{fig3} are the
coexistence pressure $p_{\rm co}$ and the corresponding packing fractions
at freezing and melting, $\eta=\eta_{\rm f}$ and $\eta=\eta_{\rm m}$,
respectively, as obtained from our MC simulations (see below). For the AO
model, the coexistence region is much broader than for the hard sphere
system. However, the freezing point is at a similar packing fraction
around $\eta\approx 0.493$ for both models. Also the fluid structure
at the freezing point is quite similar for the two systems. This can
be inferred from Fig.~\ref{fig4} where the static structure factor
\cite{hansen} $S(q)$ is displayed for the hard sphere and the AO system
at $\eta=\eta_{\rm f}$. From these findings different properties of the
solid-fluid interface may be expected for the AO system when compared
to the hard spheres. While at coexistence the fluid structure and fluid
density are very similar for both model systems, in the AO model the
fluid coexists with a fcc crystal with a much higher packing density
($\eta_{\rm m}\approx 0.64$) than in the hard sphere case where $\eta_{\rm
m}\approx 0.545$. Below we see that indeed the interfacial stiffness for
the AO system is about a factor of 2 higher than that of the hard spheres.

\begin{figure}[h]
\begin{center}
\includegraphics[width=0.4\textwidth,clip]{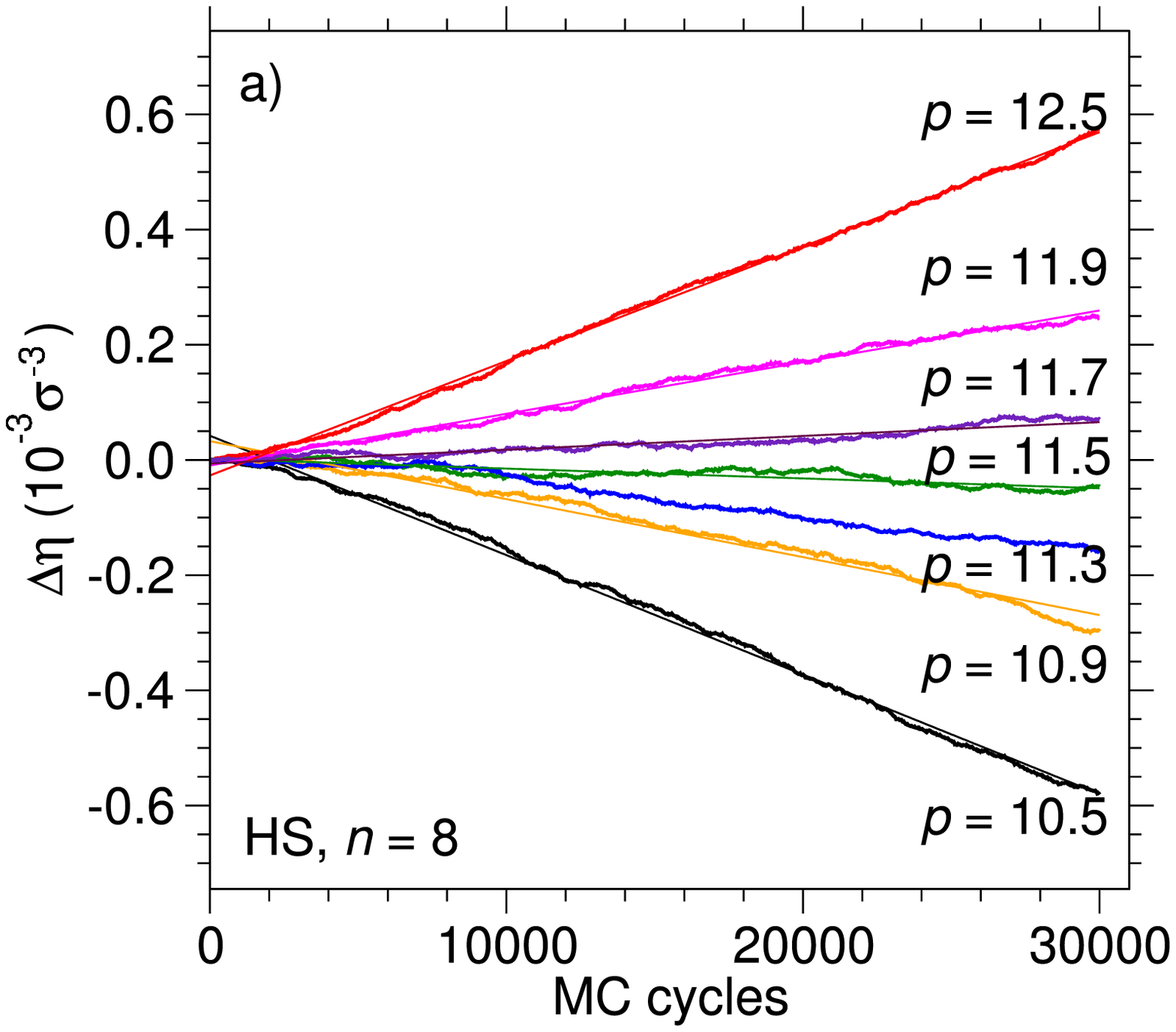}
\includegraphics[width=0.4\textwidth,clip]{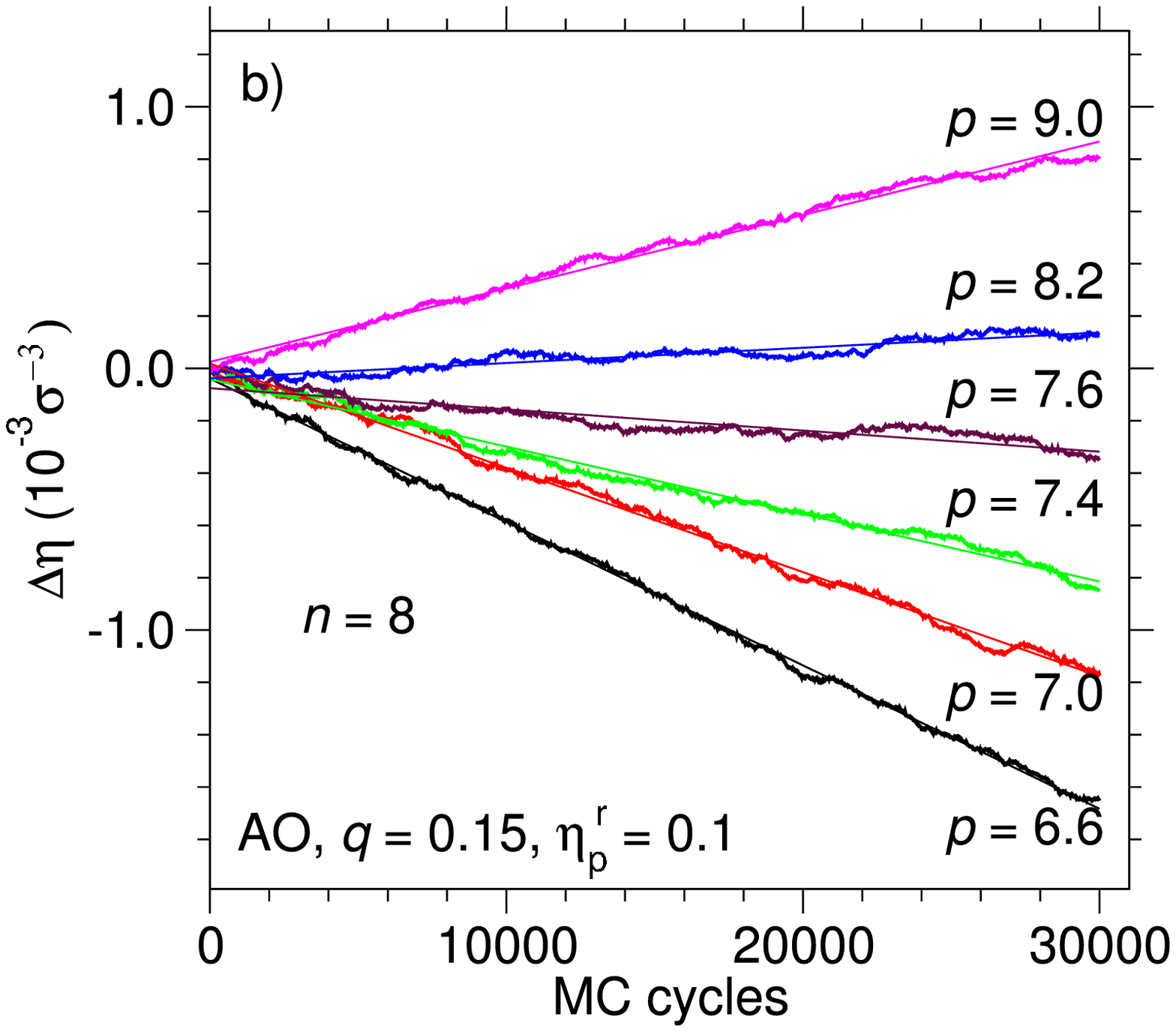}
\caption{The relative change of colloidal packing fraction, $\Delta
\eta$, as a function of Monte Carlo cycles for different pressures, as
indicated. The solid lines are fits from which $\langle
dV/dt \rangle$ is determined. a) Hard sphere system, b) AO model with
$q=0.15$ and $\eta_{\rm p}^{\rm r}=0.1$. In both cases, systems with
$N=10240$ particles are considered (corresponding to $n=8$).}
\label{fig5}
\end{center}
\end{figure}
\begin{figure}[h]
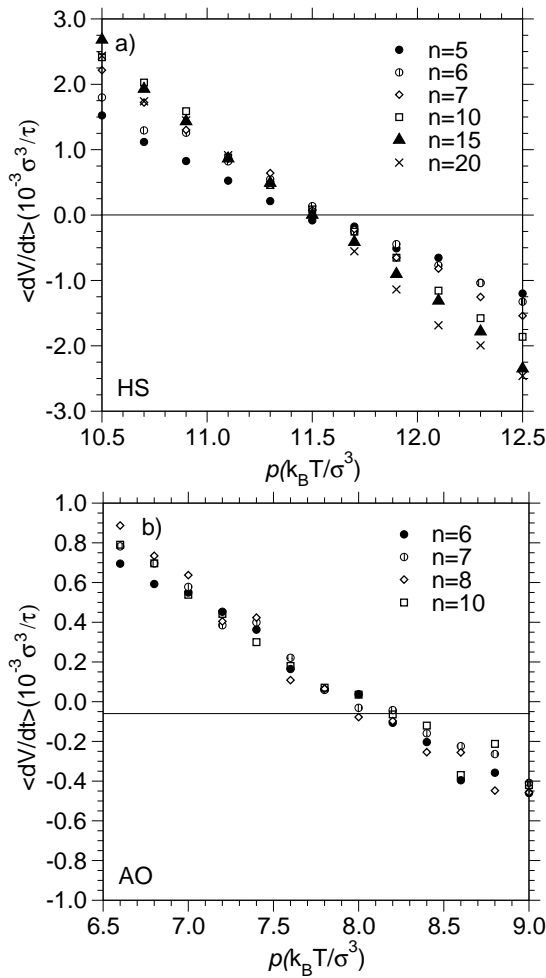

\begin{center}
\includegraphics[width=0.4\textwidth,clip]{fig6a.eps}
\includegraphics[width=0.4\textwidth,clip]{fig6b.eps}
\caption{Volume velocity $\langle dV/dt \rangle$ for different system
sizes, as indicated. a) Hard sphere system, b) AO model.}
\label{fig6}
\end{center}
\end{figure}
\begin{figure}[h]
\begin{center}
\includegraphics[width=0.4\textwidth,clip]{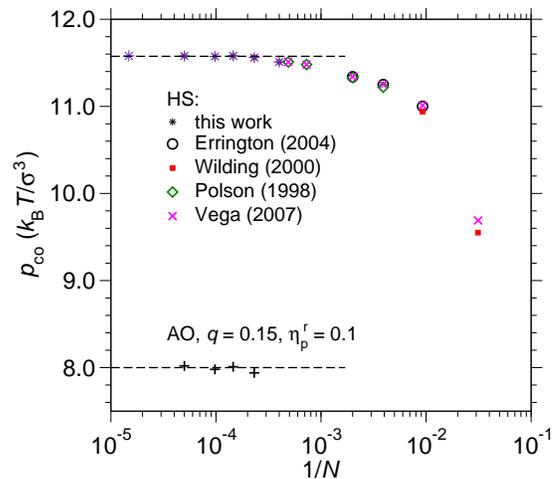}
\caption{Coexistence pressure as a function of system size (in terms
of the inverse number of particles, $1/N$) for the hard sphere system
and the AO model. The dashed lines indicate the estimated coexistence
pressures in the thermodynamic limit, $p_{\rm co}=11.576$ for the hard
sphere system and $p_{\rm co}=8.0$ for the AO model. For the hard
sphere system, results of other studies are included (from Errington
\cite{errington04}, Wilding and Bruce \cite{wilding00}, Polson and
Frenkel \cite{polson98}, and Vega and Noya \cite{vega07}).}
\label{fig7}
\end{center}
\end{figure}

Having computed the equations of state for the fluid and the solid
phases, we now switch to the simulations of the inhomogeneous systems.
As described above, the coexistence pressure $p_{\rm co}$ is obtained from
the analysis of the change of the total volume as a function of time.
The crystal grows or melts, dependent on the pressure at which the
system is considered, and thus, the coexistence point is estimated as
the state where the total volume of the system does not change with time.
Despite the simplicity of this analysis, there are several pitfalls when
applying this method. First, interactions between the interfaces due to
periodic boundary conditions have to be eliminated. Therefore, we have done
test runs with various ratios of $L_z/L$. As a result, $L_z=5L$ was found
to be an optimal choice for the avoidance of interaction effects between
the interfaces.  Second, finite-size effects need to be quantified to
obtain reliable estimates of various properties at coexistence, such as
the pressure or the interfacial stiffness. Therefore, we have considered
various system sizes with $N=2500$, 4320, 6860, 10240, 20000, 67500,
and 160000 particles.  Third, the crystal-liquid interfaces have to be
prepared such that no artificial strains are generated in the crystalline
region. In particular, the use of the isotropic $NpT$ ensemble is not
appropriate for the simulation of solid-liquid coexistence. The uniform
volume moves in $NpT$ simulations lead inevitably to strains in the $xy$
plane of the crystal, since the number of lattice planes cannot change
in these moves. Thus, the lateral dimensions of the system need to be
fixed such that they are commensurate with the chosen integer number
of lattice planes (note that the lattice spacing at a given pressure is
known from the bulk simulations of the pure fcc phase). Then, the volume is
allowed to fluctuate in $z$ direction (perpendicular to the interfaces),
keeping the pressure in that direction ($p_z$) constant.

To generate crystal-fluid samples at various pressures, first independent
solid and fluid samples were simulated.  The solid was put into a
simulation box with dimensions $L \times L \times 3L$, thereby aligning
the (100) plane of the fcc crystal perpendicular to the $z$-axis.  The box
length $L$ was chosen such that it corresponds to the solid density at
the considered pressure. At the same density, a starting configuration
for the fluid was generated by putting the particles randomly in a box
of size $L \times L \times 2L$.  Then, the box dimensions of the fluid in
$x$- and $y$-direction were kept fixed and the fluid was equilibrated by
a MC simulation in the $Np_zT$ ensemble. After $10^6$ MC cycles, solid
and fluid samples were sufficiently thermalized and put together into
a simulation box of approximate size $L \times L \times 5L$ (applying
periodic boundary conditions in all three spatial dimensions), followed
by further simulations in the $Np_zT$ ensemble. In the latter runs,
the positions of the solid particles were fixed for the first $10^5$
MC cycles to equilibrate the crystal-fluid interface without melting
away the crystal due to an unfavorable local packing of particles in
the interface region after matching the fluid slab with the solid slab.
Then, the solid particles were released for the rest of the simulation.
Finally, a set of short runs over $3\times 10^4$ MC cycles were performed
in a wide range of pressures to obtain at each pressure the total volume
as a function of time.

As an example, Fig.~\ref{fig5} shows the relative change of the colloidal
packing fraction, $\Delta \eta$, at different pressures for systems of
10240 particles. At this system size, the lateral dimension is given by
$L=L_x=L_y=na$ with $n=8$ lattice planes in units of the lattice constant
$a$ (of course, $a$ changes as a function of pressure both for the AO
and the HS system).  In the following, we indicate the system size in
terms of the number $n$, considering box geometries of nominal size $L
\times L \times 5L$.  We also note that the time $t$ is measured in units
of the number of Monte Carlo cycles; of course, it cannot be directly
translated into a physical time, but, since we are not interested in
the growth kinetics here, this does not matter in the present context.

The slopes $\langle dV/dt \rangle$, averaged over 10 independent
configurations, are displayed in Fig.~\ref{fig6} for different system
sizes, ranging from $n=5$ to $n=20$ for the hard sphere system and from
$n=6$ to $n=10$ for the AO system. The values for $\pco$, as estimated
via the interpolation to $\langle dV/dT \rangle = 0$ for the different
system sizes, are plotted in Fig.~\ref{fig7} as a function of the inverse
number of particles $1/N$.  Obviously, both for the HS and the AO model,
finite-size effects are small in the considered range of system sizes and
we obtain $\pco = 11.576 \pm 0.006 \un$ for the hard spheres and $\pco =
8.0 \pm 0.026 \un$ for the AO model.  The corresponding packing fractions
for freezing and melting are respectively given by $\eta_{\rm f}=0.492$
and $\eta_{\rm m}=0.545$ for the hard spheres and by $\eta_{\rm f}=0.494$
and $\eta_{\rm m}=0.64$ for the AO model (see also Figs.~\ref{fig2} and
\ref{fig3}).

Also included in Fig.~\ref{fig7} are estimates of $\pco$ for
the hard sphere system, as obtained from other simulation studies
\cite{errington04,wilding00,polson98,vega07}.  In these studies, the use
of thermodynamic integration techniques as well as the phase switch MC
method allowed only the consideration of relatively small system sizes
and so these results lie below the dashed line in Fig.~\ref{fig7} that
marks the estimate of $\pco$ in the thermodynamic limit, as obtained from
our simulation. This indicates the advantage of the methodology used in
this work: relatively large system sizes can be simulated and thus it
is not necessary to perform extrapolations to the thermodynamic limit
(at least not for the systems considered here) that may easily lead to
systematic errors in the estimate of the coexistence pressure.

\subsection{Local order parameters and interfacial structure}
\label{sec:order}
\begin{figure}[h]
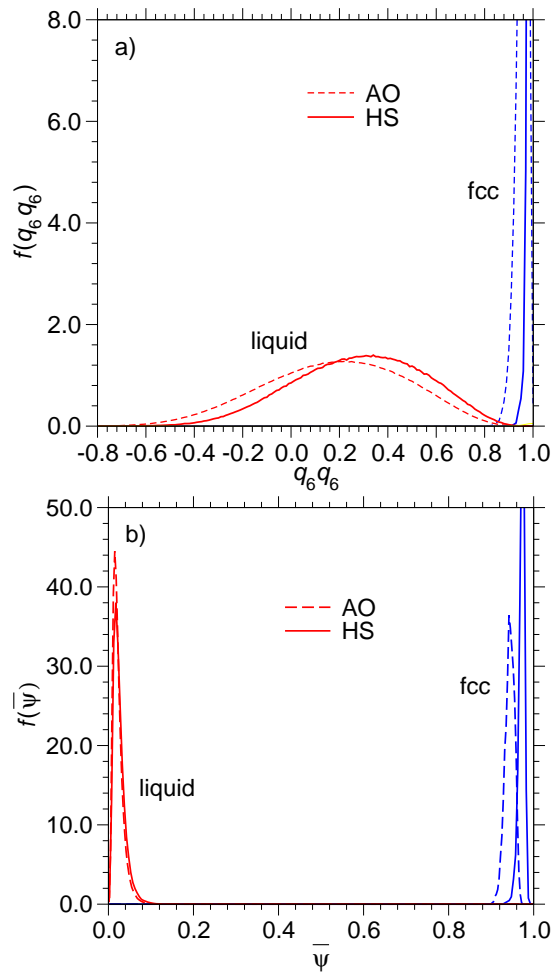

\begin{center}
\includegraphics[width=0.4\textwidth,clip]{fig8a.eps}
\includegraphics[width=0.4\textwidth,clip]{fig8b.eps}
\caption{Order parameter distributions for the AO model (dashed lines)
and the hard sphere system (solid lines) for the liquid and the fcc phase,
as indicated; a) $q_6q_6$, b) $\bar{\Psi}$.}
\label{fig8}
\end{center}
\end{figure}
\begin{figure}[h]
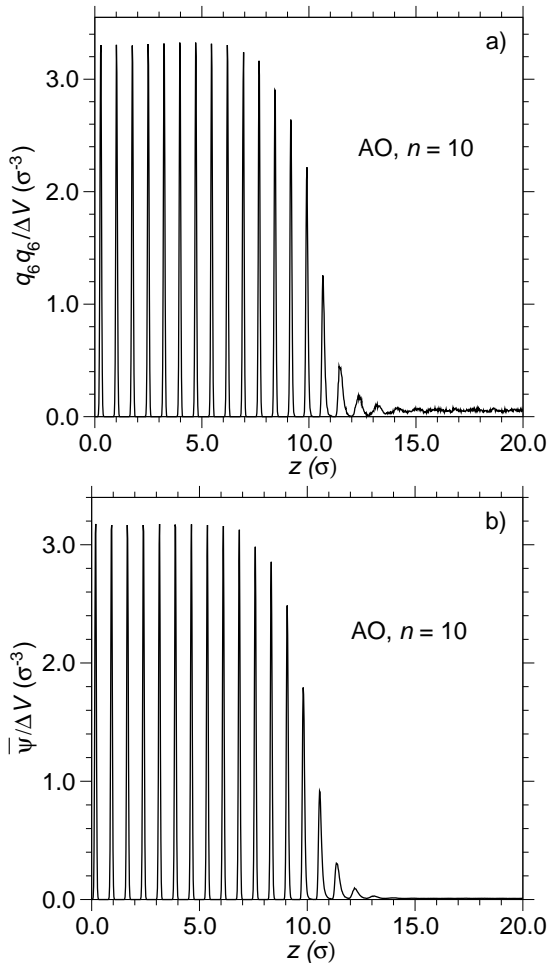

\begin{center}
\includegraphics[width=0.4\textwidth,clip]{fig9a.eps}
\includegraphics[width=0.4\textwidth,clip]{fig9b.eps}
\caption{Fine-grained order parameter profiles of a) $q_6q_6$ and
b) $\bar{\Psi}$ for the AO model. The system contains 20000 particles,
corresponding to $n=10$.}
\label{fig9}
\end{center}
\end{figure}
Having determined $\pco$ for the hard sphere and the AO model, we can
now investigate properties of solid-fluid interfaces in these systems
at coexistence.  First, order parameters have to be identified that
allow to distinguish between solid-like and fluid-like local order
around a particle. A class of order parameters that is well-suited for
this purpose are local bond order parameters, $Q_l(i)$, as introduced
by Steinhardt {\it et al.}~\cite{steinhardt83}. They are based on the
expansion into spherical harmonics $Y_{lm}$:
\begin{equation}
\label{eq:localorder2}
Q_l(i)=
\left(
\frac{4\pi}{2l+1}\sum_{m=-l}^{l}{|\bar{Q}_{lm}|^2}
\right)^{1/2}
\end{equation}
with
\begin{equation}
\label{eq:localorder1}
\bar{Q}_{lm}(i)=
\frac{1}{Z_i}
\sum_{j=1}^{Z_i}{Y_{lm}(\theta(\vec{r}_{ij}),\phi(\vec{r}_{ij}))} \, ,
\end{equation}
where $\vec{r}_{ij}$ is the distance vector between a pair of neighboring
particles $i$ and $j$, $Z_i$ is the number of neighbors within a given
cut-off radius, and $\theta(\vec{r}_{ij})$ and $\phi(\vec{r}_{ij})$
are the polar bond angles with respect to an arbitrary reference frame.

A variant of the order parameters (\ref{eq:localorder2}) has been put
forward by ten Wolde {\it et al.}~\cite{tenwolde95} by introducing the
dot product
\begin{equation}
\label{eq:localorder3}
q_lq_l(i)=
\frac{1}{Z_i}\sum_{j=1}^{Z_i}
\sum_{m=-l}^{l}{\tilde{q}_{lm}(i)\tilde{q}_{lm}(j)^*} \, ,
\end{equation}
with
\begin{equation}
\label{eq:localorder5}
\tilde{q}_{lm}(i)=
\frac{\bar{Q}_{lm}(i)}
{\left(\sum_{m=-l}^{l}{|\bar{Q}_{lm}(i)|^2}\right)^{1/2}} \, .
\end{equation}
Here, we use $q_6q_6$ that is defined by Eqs.~(\ref{eq:localorder3})
and (\ref{eq:localorder5}), setting $l=6$.

Figure \ref{fig8}a shows the $q_6q_6$ distributions for the pure fluid
and fcc phases at coexistence for the hard sphere system (solid lines)
and the AO mixture (dashed lines). The relatively small overlap of the
distribution for the solid with that of the fluid indicates that $q_6q_6$
is well-suited to distinguish between solid and fluid particles. Note
that we have used time-averaged particle positions for the calculation
of the order parameter distributions in Fig.~\ref{fig8}a (also for
the distributions shown in Fig.~\ref{fig8}b).  Particle positions were
averaged over 50 MC cycles in case of the hard sphere system and over 20
MC cycles in case of the AO mixture. This reduces the shift of the order
parameter distributions for the solid fcc phases to lower values of the
order parameter, as compared to the distribution of an ideal fcc crystal.

A different local order parameter has been introduced by
Morris~\cite{morris02_1,morris02}. It is defined by
\begin{equation}
\label{eq:localorder6}
\Psi(i)={\Big |} \frac{1}{N_q}\frac{1}{Z_i}
\sum_{j=1}^{Z_i}\sum_{k=1}^{N_q}
\exp(i\vec{q}_k\cdot \vec{r}_{ij}) {\Big |}^2
\end{equation}
where $\vec{r}_{ij}$ denotes the distance vector of a particle $i$
to neighboring particles $j$, and the wave-vectors $\vec{q}_k$ are
related to reciprocal vectors of the fcc lattice with lattice constant
$a_0$, $\vec{q}_1=2\pi/a_0(-1,1,-1)$, $\vec{q}_2=2\pi/a_0(1,-1,1)$
and $\vec{q}_3=2\pi/a_0(1,1,-1)$.  An additional average of $\Psi(i)$
over a particle with index $i$ and its neighboring particles in the
first coordination shell yields
\begin{equation}
\label{eq:localorder8}
\bar{\Psi}(i)=
\frac{1}{Z_i+1}\left(\Psi(i)+\sum_{j=1}^{Z_i}\Psi(j)\right)
\end{equation}
As can be inferred from Fig.~\ref{fig8}b, the $\bar\Psi(i)$ distributions
display a sharp peak close to $\bar{\Psi}=1$ for the fcc phase and one
close to $\bar{\Psi}=0$ for the fluid. Obviously, the order parameter
$\bar{\Psi}$ is also well-suited to identify the local order around
particles.

By applying the methodology of our previous study~\cite{tanya09}, we
characterize the local structure of the interfaces by $z$-dependent
profiles of averaged local order parameters. For this purpose
solid-fluid samples were divided into bins of length $\Delta z =0.05 \sigma$
along the $z$-direction and ''time`` averages of the order parameter
were obtained for each bin, thereby correcting for shifts of the crystal
planes with respect to the reference frame during the simulation.

Figures \ref{fig9}a and \ref{fig9}b show order parameter profiles for
the AO model as a function of the $z$-coordinate (i.e.~perpendicular to
the interface). To obtain these profiles, the sum of the order parameter
of the particles in each bin transversal to the solid-fluid interface
is divided by the volume of the bin $\Delta V = L^2 \Delta z$.  In the
crystalline region, the profiles exhibit strong oscillations with the
periodicity of the crystalline planes. The amplitude of these oscillations
decay rapidly in the interfacial region and flatten completely in the
fluid region.  Corresponding data for the hard-sphere system are reported
in our previous study~\cite{tanya09}.

\subsection{Finite-size interfacial broadening and interfacial stiffness}
\label{sec:cwm}
\begin{figure}[h]
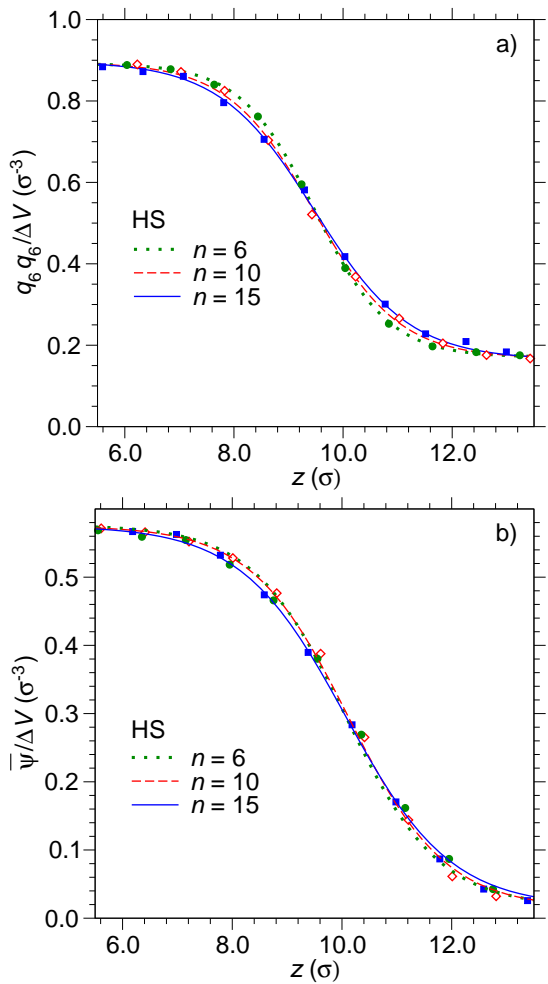

\begin{center}
\includegraphics[width=0.4\textwidth,clip]{fig10a.eps}
\includegraphics[width=0.4\textwidth,clip]{fig10b.eps}
\caption{Coarse-grained order parameter profiles for the hard sphere
system for different system sizes; a) $q_6q_6$, b) $\bar{\Psi}$. The
lines are fits to Eq.~(\ref{eq_tanh}).}
\label{fig10}
\end{center}
\end{figure}
\begin{figure}[h]
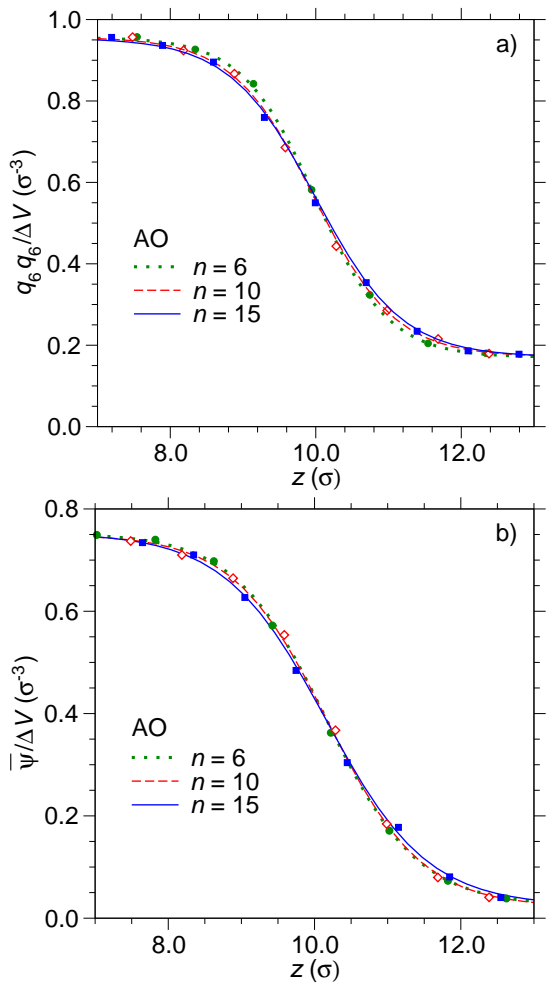

\begin{center}
\includegraphics[width=0.4\textwidth,clip]{fig11a.eps}
\includegraphics[width=0.4\textwidth,clip]{fig11b.eps}
\caption{Coarse-grained order parameter profiles for the AO model for
different system sizes; a) $q_6q_6$, b) $\bar{\Psi}$. The lines are fits
to Eq.~(\ref{eq_tanh}).}
\label{fig11}
\end{center}
\end{figure}
Now, we aim at computing the interfacial stiffness $\tilde{\gamma}$
which is defined by $\tilde{\gamma}= \gamma + d^2 \gamma/d\theta^2$
with $\gamma$ the interfacial tension and $\theta$ the angle between the
interface normal and the (100) direction. Whereas $\gamma$ describes
the free energy cost to increase the area of the interface, the term
$d^2 \gamma/d\theta^2$ accounts for the free energy required to locally
change the orientation of the crystal. The latter term would of course
vanish for a system where both phases at the interface are isotropic
(as, e.g., in the case of liquid-gas interfaces).

In the framework of capillary wave theory (CWT), the interfacial stiffness
can be computed from the broadening of the solid-fluid interface with
increasing lateral size of the system \cite{tanya09} (see below). To
make use of this prediction, the apparent mean-squared width of the
interface, $w^2$, has to be determined and analyzed as a function of
system size. For this purpose, fine-grid order profiles as the ones shown
in Fig.~\ref{fig9} are not well-suited since the oscillations due to the
crystalline order do not allow to fit the profile with a simple function
where the interfacial width appears as a free parameter. Therefore, it is
useful to coarse-grain the profiles by averaging over the oscillations. To
this end, we have identified the minima in the fine-grid profiles and used
them to mark the borders of non-uniform bins that match the crystalline
layers. Then, we took the average of the order parameter in each of the
latter bins.

Examples of the resulting coarse-grained profile for different system
sizes are shown in Figs.~\ref{fig10} and \ref{fig11}. As we can see in
these plots, the data can be well fitted with hyperbolic tangent function
of the following form
\begin{equation}
\label{eq_tanh}
\phi(z) = \frac{A+B}{2} + \frac{A-B}{2}\tanh\left(\frac{z-z_0}{w}\right)
\end{equation}
where $A$ and $B$ are the bulk order parameters in the solid and fluid
obtained from the independent bulk simulations. So the position of the
interface $z_0$ and its width $w$ are the only fit parameters.  Note that
Eq.~(\ref{eq_tanh}) can be obtained in the framework of mean-field
theory. However, here it is just used a fitting function to determine
the width $w$ as a function of the lateral size of the system. Figures
\ref{fig10} and \ref{fig11} indicate that both for the hard sphere model
and the AO mixture, $w$ slightly increases with increasing system size.

\begin{figure}[h]
\begin{center}
\includegraphics[width=0.35\textwidth,clip]{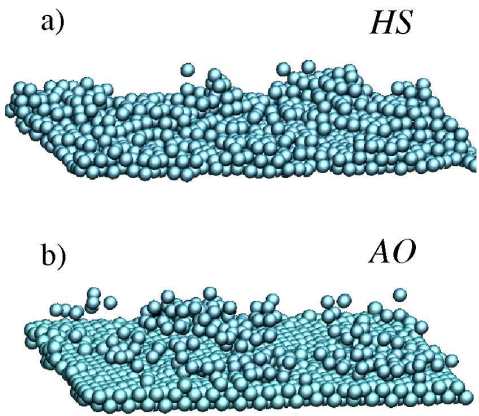}
\caption{Snapshots of the interfaces for a) hard spheres and b) the AO
model for systems with $n=15$. Here, only solid particles are shown,
defining a particle as a solid one when its $q_6q_6 > 0.6$ 
and its coordination number greater than 10.}
\label{fig12}
\end{center}
\end{figure}
\begin{figure}[h]
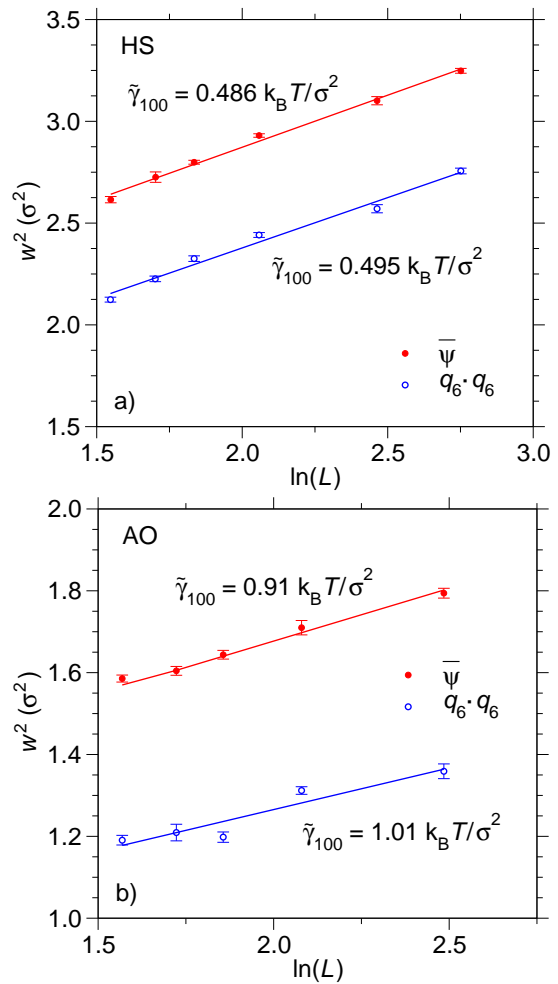

\begin{center}
\includegraphics[width=0.4\textwidth,clip]{fig13a}
\includegraphics[width=0.4\textwidth,clip]{fig13b}
\caption{Mean-squared width $w^2$ as a function of $\ln L$, for a)
hard spheres and b) the AO model. The values for $\tilde{\gamma}$
are obtained from the fits (solid lines) to Eq.~(\ref{eq_width}).}
\label{fig13}
\end{center}
\end{figure}
The finding that the width $w$ is system-size-dependent is expected
from CWT where the mean-squared interfacial width is found to
diverge logarithmically with the lateral size $L$ of the system
\cite{buff65,weeks77,bedeaux85,gelfand90},
\begin{equation}
\label{eq_width}
 w^2 = w^2_0 + \frac{k_BT}{4\tilde{\gamma}} {\rm ln} \frac{L}{l} \, ,
\end{equation}
with $w_0$ the so-called intrinsic width and $l$ a cut-off parameter
that has to be introduced since CWT only describes long-wavelength
undulations of the interface.  Note that it is impossible to disentangle
the intrinsic width contribution $w_0^2$ from the cut-off contribution
$\frac{k_BT}{4\tilde{\gamma}} {\rm ln} l$. An extensive discussion
of this issue has been provided for the case of polymer mixtures
\cite{werner97,werner99_1,werner99_2,kerle99,binder00}.

The existence of capillary waves is associated with the
roughness of the interface and the applicability of CWT
requires that interfaces above a possible roughening transition
\cite{shugard78,burkner83,mon88,mon89,schmid92} are considered.  That both
for the hard sphere system and the AO mixture the solid-fluid interface
is indeed rough, is indicated by the snapshots in Fig.~\ref{fig12}.
Here, only the solid particles are shown, defining a particle as a solid
one if $q_6q_6>0.6$ and if its nearest-neighbor coordination number is
larger than 10 (note that we have only needed this definition for the
snapshots in Fig.~\ref{fig12}). The snapshots reveal that the solid-fluid
interface of the hard sphere system is significantly ``more rough''
than that of the AO mixture.  This might be due to the larger density
difference between the fcc and the fluid phase in case of the AO model,
when compared to the hard sphere system.

In Fig.~\ref{fig13}, the prediction (\ref{eq_width}) is confirmed:
both for the hard spheres and the AO mixture, the mean-squared
width $w^2$ exhibits a logarithmic divergence with respect to the
lateral dimension $L$. Within the statistical errors, the two order
parameters $q_6q_6$ and $\bar{\Psi}$ yield similar results.  From the
fits with Eq.~(\ref{eq_width}), we estimate for the (100) orientation
$\tilde{\gamma} \approx 0.49\pm 0.02$\,k$_{\rm B}T/\sigma^2$ for the
hard sphere system and $\tilde{\gamma} \approx 0.95\pm 0.1$\,k$_{\rm
B}T/\sigma^2$ for the AO mixture. The value of $\tilde{\gamma}$
for the hard sphere system roughly agrees with previous estimates,
obtained by other methods (for a discussion of this issue, see
Refs.~\cite{laird05,tanya09}).

\section{Conclusions}
\label{sec:conclusions}
We have investigated the fluid-to-crystal transition of two paradigmatic
colloidal model systems using Monte Carlo (MC) simulations at constant
pressure.  Inhomogeneous systems have been prepared where the crystal
phase in the middle of an elongated simulation box is separated from the
fluid phase by two interfaces. We have demonstrated that MC simulations
with this setup allow for reliable estimates of the coexistence pressure,
$\pco$, and the interfacial stiffness, $\tilde{\gamma}$. Both for the
hard sphere system and the AO mixture, our methodology allows for the
simulation of relatively large systems and thus, the coexistence pressure
$\pco$ can be computed without relying on error-prone extrapolations
to the thermodynamic limit.  On the other hand, we have presented
a method for the calculation of $\tilde{\gamma}$ that makes use
of finite-size effects due to the capillary-wave broadening of the
interface. So far, this method has been mainly applied to polymer
interfaces \cite{werner99_1}, liquid-vapor interfaces \cite{vink05},
isotropic-nematic interfaces \cite{wolfsheimer06} or interfaces in
Ising systems \cite{mueller05}, but not to solid-fluid interfaces,
as in this work (an exception is of course our recent preliminary work
on hard spheres and nickel \cite{tanya09}).  It requires a systematic
variation of the system size and relatively small statistical errors
to resolve the logarithmic divergence of the mean-squared interfacial
width, as predicted in the framework of capillary wave theory (CWT).
With respect to the latter issues, the determination of $\tilde{\gamma}$
is much more difficult for the AO mixture than for the hard sphere system.
First, in the relevant range of colloid packing fractions $\eta$ mass
transport processes of the AO fluid in the bulk and in the interface
region are much slower than in the hard sphere system. Therefore, much longer
simulation runs are necessary for the AO model, to achieve comparable
statistics as in the hard sphere system. Second, the interfacial stiffness
$\tilde{\gamma}$ for the AO model is about a factor of two higher than
in the hard sphere system, associated with lower-amplitude capillary
fluctuations at comparable system sizes.  Thus, the signal-to-noise ratio
is worse for the AO model when one analyzes the interfacial fluctuations.
However, despite the latter difficulties, the estimate of $\tilde{\gamma}$
from the interfacial broadening works also well for the AO model.

CWT can be of course only applied to the analysis of interfacial
properties, if rough, non-faceted interfaces are considered.  Otherwise,
there would be no divergence of the width of the interface with increasing
lateral size of the system. For the case of the model systems considered
in this work, the applicability of CWT is justified since we observe the
logarithmic growth of the mean-squared interfacial width as a function
of the lateral dimension $L$.  To further rationalize the use of CWT,
we plan to study the Fourier spectrum of interfacial fluctuations from
which one can alternatively determine the interfacial stiffness. Moreover,
we plan to investigate the possibility of a roughening transition of
the AO model where the solid-fluid interface changes from a faceted to
a rough interface with long-wavelength capillary wave fluctuations.

\begin{acknowledgments}
We thank Philipp Kuhn for a critical reading of the manuscript.
We are grateful to the German Science Foundation (DFG) for financial
support in the framework of the focus program SPP 1296.  We acknowledge
a substantial grant of computer time at the SOFTCOMP of the John von
Neumann Institute for Computing (NIC) and the computer center (ZDV)
of the Johannes Gutenberg-Universit\"at Mainz.
\end{acknowledgments}

\end{document}